\documentclass{svmult}

\usepackage{url}
\usepackage{makeidx}         % allows index generation
\usepackage{graphicx}        % standard LaTeX graphics tool
\usepackage{multicol}        % used for the two-column index
\usepackage[bottom]{footmisc}% places footnotes at page bottom
\usepackage[format=plain, font=small, labelfont={sf,bf}]{caption, subfig}
\usepackage[latin1]{inputenc}
\usepackage{amsmath}
\usepackage{amssymb}                    % Schriftarten der A.M.S
\usepackage{color}
\usepackage{bm}

\usepackage{booktabs}           % Buchtabellen
\usepackage{placeins}		% \FloatBarrier

\usepackage{tikz}   			% zum Anzeigen von gnuplot Diagrammen
\usetikzlibrary{positioning}

\usepackage{pgfplots}
\pgfplotsset{compat=newest}
\usepgfplotslibrary{colormaps}

\usetikzlibrary{plotmarks}
\usetikzlibrary{arrows.meta}
\usetikzlibrary{patterns} %for hatched area in CellList.tex
\usepgfplotslibrary{patchplots}
\usetikzlibrary{pgfplots.fillbetween}
\usetikzlibrary{matrix,calc}
\usetikzlibrary{shapes,arrows}

\newcommand{\vek}[1]{\mathbf{#1}} %kursiv Fett
\begin{document}

\title*{Advanced Boundary Conditions for the Simulations of Laser
  Ablation}
\titlerunning{Boundary Conditions for Laser Ablation}

\author{Eugen Eisfeld, Dominic Klein, Johannes Roth}
\authorrunning{E. Eisfeld, D. Klein, J. Roth}
    \institute{
    Eugen Eisfeld \at Institut f\"ur Funktionelle Materie und
    Quantentechnologien, Universit\"at Stuttgart,
    \email{eugen.eisfeld@fmq.uni-stuttgart.de}
    \and
    Dominic Klein \at Institut f\"ur Funktionelle Materie und
    Quantentechnologien, Universit\"at Stuttgart,
    \email{dominic.klein@fmq.uni-stuttgart.de}
    \and
    Johannes Roth \at Institut f\"ur Funktionelle Materie und
    Quantentechnologien, Universit\"at Stuttgart,
    \email{johannes.roth@fmq.uni-stuttgart.de}
    }
%
% Use the package "url.sty" to avoid
% problems with special characters
% used in your e-mail or web address
%
\maketitle

\abstract{Irradiating materials with ultra-short laser pulses generates
  unwnated shock waves which distort the interesting physics. Advanced
  boundary conditions are presented to erase even very strong shock
  waves. Depending on the wave length, laser light which leads to ablation may
  penetrate very deep into the sample. Connection conditions are presented to
  consistently couple atomistics and continuum description of the sample
  even at high temperatures together with a discussion of the temperature
  definition.}

\section{Introduction}

This report will deal with some very important aspects of the atomistic
molecular dynamics (MD) simulations of laser ablation coupled with the
so-called two-temperature-model (TTM) for the continuum treatment of
electrons. Details of the two model parts have been given in a previous HLRS
report\cite{md-ttm} and by Schaefer and Urbassek \cite{Schafer2002} for
example, and will not be repeated  here due to the lack of space. Meanwhile,
the physical MD+TTM model has been improved considerably for metals in the
thesis of Eisfeld \cite{eisfeld2020}  and for covalent materials in the work
of Klein \cite{klein2021}.

Even on the biggest available supercomputers atomistic computer
simulations are limited to a few micrometers due to the storage requirements
of all the atoms and to a few microseconds due to the integration steps of
femtoseconds if a typical maximum of a million of stepps is assumed.  

There are three aspects which can improve the yield of large scale
computing considerably: let the first improvement be a better (physical) model
and the second a better implementation. Then the third aspect is the reduction
of the simulated sample size by special boundary conditions which account for
long-range effects. In the case of laser ablation the problem
has actually two facets: first of all, shock and tension waves are generated
which lead to unwanted effects influencing the ablation process. More
generally these processes are directed energy transmissions which do not play
a role in macroscopic experiments since there they die out, but are reflected
at boundaries in atomistic simulations and pile up, thereby disrupting the 
simulation sample for example. The second facet is the deep penetration of
light into material which requires long samples to simulate time evolution. As
the original two-temperature model was formulated as a continuum model not
only for the electrons but for the atoms also, it is self-evident to switch
back to this model deep in the sample. What is required, are connection
conditions between the atomistic part and the continuum part.
The paper is organized as follows. After a short digest of heat absorbing
boundary conditions for completeness, pressure absorbing boundary conditions
are introduced to solve the shock wave problem. Then the spatial coupling of the
atomistic part with the continuum description is described together with a
discussion of the temperature definition which is necessary for a consistent
cooperation of both parts.

\section{Heat absorbing boundary conditions}

In atomistic materials science simulations samples are often heavily deformed
such that they show phase transformations, or cracks are generated for
example. The consequence are breaking bonds releasing energy at the crack
surface. The sample heats up and uncontrollable melting of the whole sample
occurs. In macroscopic experiments this is not a 
problem since the samples are large enough such that the heat can be
compensated. Cracks emit dispersive heat waves which can easily be absorbed by
a so-called ``stadium'' surrounding the crack \cite{gumbsch}. Since the heat
is spread in all directions, the methods work quite well and are well 
established. In one approach the equations of motion outside the stadium are
supplemented with a friction term which removes the heat. In another approach
a NVE ensemble which conserves energy is simulated within the stadium and a
NVT ensemble is implemented outside the stadium effectively cooling the sample
to the desired temperature and thus removing the released energy.

\section{Pressure absorbing boundary conditions}\label{subsec:NRB}
An important problem which has to be addressed, regarding the economical
simulation of laser ablation is the reflection of shock waves. 
Such shock waves move at the speed of sound and beyond through the
sample and are reflected at the rear surface of the sample. As soon as the
reflected wave reaches the original sample surface again it may influence the
alation process considerably, distort the behavior and change the results. In
general, this can be avoided by simulating samples long enough, such that the
reflected wave arrives at the original surface only when the ablation process
hs already completed. Thereforce, a considerable fraction of processing power
and memory is wasted only to to compensate for this undesirable effect. 
This can be avoided by applying pressure absorbing boundary conditions.

A linear or quadratically increasing ``damping ramp'' \cite{ulrich, DissSonntag}
which extracts gradually the kinetic energy of the atoms represents a big 
improvement already. 
In this case however, the extent of the damping ramp can not be chosen arbitrarily short. 
The friction term may only increase very slowly otherwise an artificial impedance between neighbouring regions is introduced, 
resulting again in a reflection.
Moreover, an energy ramp also represents a considerable heat sink which, in case of a
minimally chosen ramp length may unnaturally increase the heat conduction within
the heat-affected zone by amplifying the temperature gradient. 

The supposedly most efficient technique for absorption of sound and shock
waves in atomistic simulations is based on the so-called
``time-history-kernel'' method\cite{pang2011,Park2005,Karpov2006}. 
In this method the equations of motion of the atoms at the free surface are
modified as if they were surrounded by neighboring atoms from all sides. 
The major drawback of this method lies in the fact that it requires
the forces of all neighboring atoms for all previous time steps to be present in the current memory, 
since the modification of the equations of motion rests on a convolution of
the atomic trajectories. 
A considerable simplification is provided by Schaefer and
Urbassek\cite{Schafer2002}. 
They also modify the equations of motion of the boundary atoms, however, only one restoring force depending on the momenta of the
boundary atoms and their center of mass drift velocity is added to the original forces.
The latter method has been tested elaborately, however, the success of this method, as 
mentioned in the publication could be reproduced only for relatively weak shock waves. 

A considerable improvement is provided by the absorbing boundary conditions 
according to Ming and Fang\cite{ming2009,Fang2012}. 
By additionally considering the momenta and positions of the next neighbors the authors formulated
on the basis of a simple spring-sphere model with linearized forces a highly
efficient solution to the problem. 
Furthermore they abstained from the simplification of the shock wave as a scalar wave and treated the involved atomistic
displacements in a complete vector wave picture. 
In our test simulations, their modified equations of motion for the boundary atoms withstood shock waves with
intensities up to 40~GPa and beyond, as long as the material in close
proximity to the free surface did not lose its original crystallinity due
to phase transitions or plastic deformation.

\begin{figure}
 \centering
 \includegraphics[width=0.7\textwidth]{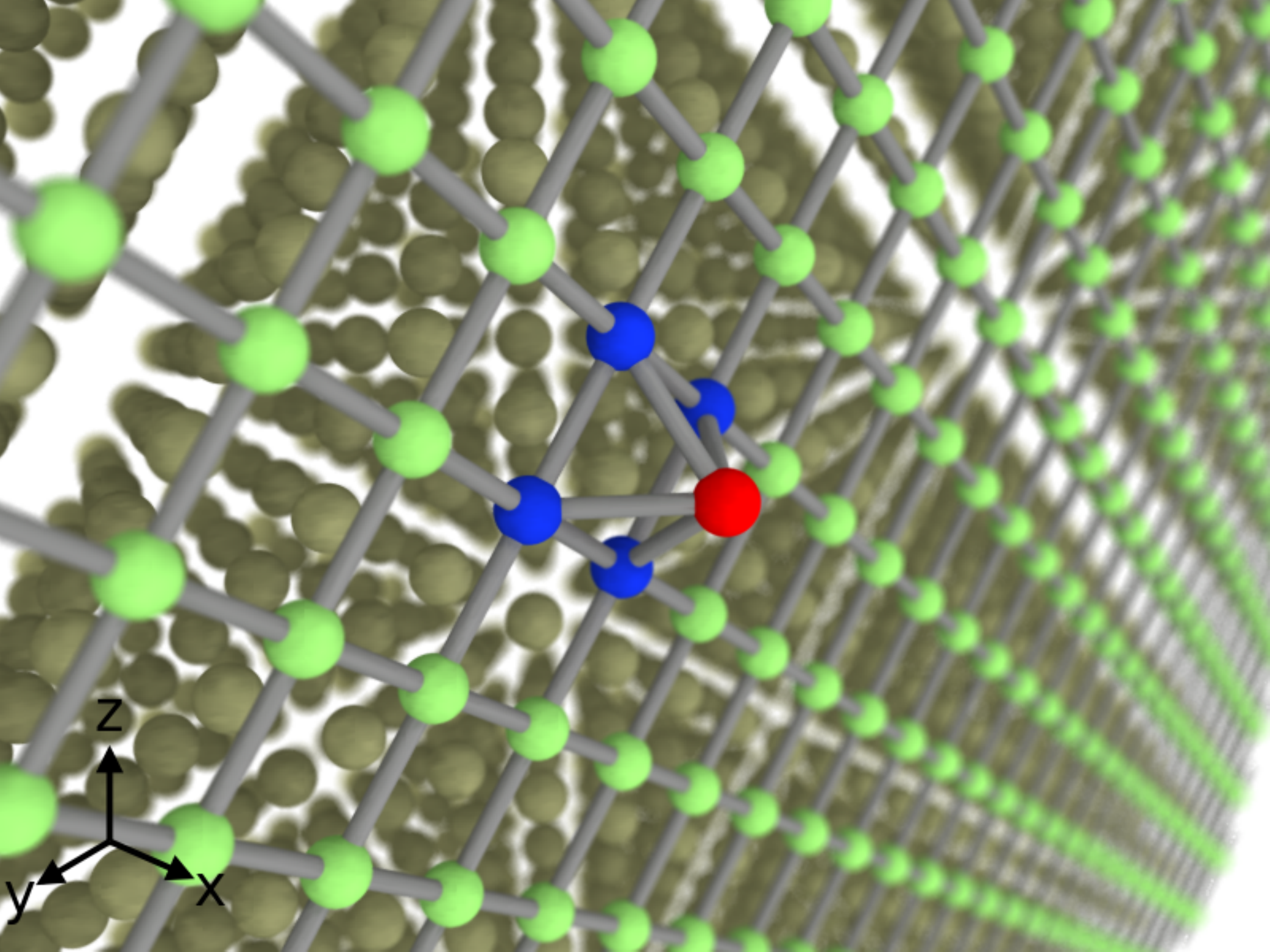}
 \caption{Perspective representation of the nearest neighbor atoms (blue) of
   a selected boundary atom (red) in a (100)-oriented cubic face-centered
   crystal.}
  \label{fig:nrbpersp}
\end{figure}

A more elaborate description of the model with appropriate derivations may be
found in the original publications. 
Here we restrict ourselves to the formulation of the modified equations of
motion for the boundary atoms in a (100)-oriented face centered cubic crystal (fcc)\footnote{Where the crystallographic $(1,0,0)^T$-direction is parallel to the simulation box axis $x$.}. 
The positions of the atoms are given by the indices $l,m,n
\in \mathbb{Z}_0$ of the lattice sites. 
For the boundary atoms, namely the atoms belonging to the lattice plane at the free surface, $l=0$. 
For the neighboring atoms a lattice plane deeper into the crystal $l=1$. 
The displacement $\vek{u}_{0,m,n}$ of a boundary atom from is original equilibrium position
follows the equation of motion
\begin{align}
&\vek{\dot{u}}_{0,n,m}=\sqrt{\frac{k}{M}}\vek{u}_{0,n,m}\cdot
\begin{pmatrix}
1 &  0 & 0\\
0 & -2\sqrt{2} & 0\\
0 &  0 &-2\sqrt{2}
\end{pmatrix}\notag\\
&+
\sqrt{\frac{k}{M}}\vek{u}_{0,n,m}
\begin{pmatrix}
1 & 0 & 0\notag\\
0 & \sqrt{2}/2 & 0\\
0 & 0 & \sqrt{2}/2
\end{pmatrix}
\left(\vek{u}_{1,n,m+1} + \vek{u}_{1,n,m-1}+\vek{u}_{1,n+1,m}+\vek{u}_{1,n-1,m}\right)\\
&-\frac{1}{4}%
\left(\vek{\dot{u}}_{1,n,m+1} +
\vek{\dot{u}}_{1,n,m-1}+\vek{\dot{u}}_{1,n+1,m}+\vek{\dot{u}}_{1,n-1,m}\right)\,, 
\end{align}
where $k$ is the force constant and $M$ the mass of the atom according to the
sphere-spring model. Thus, only the four nearest neighbors in the lattice plane
$l=1$ have to be taken into account. This situation is represented in Fig.~\ref{fig:nrbpersp}. 

The force constant can be computed from the potential
analogously to the elaboration of Schaefer and Urbassek\cite{Schafer2002}. 
Alternatively, it can be fitted to the particular intensities of the shock waves in a series of test
simulations. 
The latter procedure is more tedious, but in the end leads to
better results since here, effects which go beyond the strongly simplified
sphere-spring model can also be compensated for. 

\begin{figure}
\centering
\input{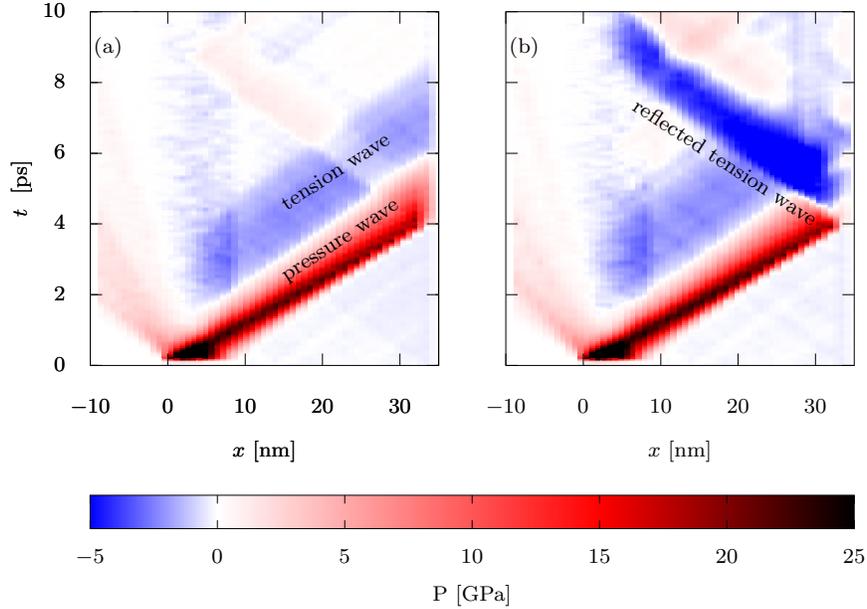}
\caption{Histogram plot of the temporal pressure-evolution after sudden heating of
  the surface of a (100)-oriented aluminum single crystal. (a) represents
  the situation with pressure absorbing boundary conditions, 
  whereas (b) demonstrates what happens without these modified equations of motion. 
  Along the $x$-axis the position below the original surface $x=0$ is displayed.}  
\label{fig:nrbpressplot}
\end{figure}

For our model material aluminum, satisfactory results could be achieved with
$k=0,9\,\mathrm{eV/\mathring{A}^2}$. 
The efficiency of the method can be seen in the contour plots for 
the temporal evolution of the hydrostatic pressure \ref{fig:nrbpressplot}. 
The sample surface in this quite extreme example is heated during the first 300~fs so strongly, 
that the generated shock wave hits the rear side of the sample, which is only a few nanometers apart from the original surface with an intensity of about 25~GPa. 
Without pressure absorbing boundary conditions the wave is reflected at the rear side and its sign is flipped, 
giving rise to an intensity of 5~GPa. 
As a consequence, the rear side of the sample bursts into numerous fragments, a mechanism commonly referred to as spallation.
In addition, the reflected wave reaches the front surface shortly thereafter. 
The application of the pressure absorbing boundary conditions on the contrary only leads to a temporary buckling at the
rear side of the sample. 
The crystal structure remains intact and spallation is suppressed. 
Certainly, the reflection in this extreme case cannot be avoided
completely but there is no doubt that the intensity of the wave is largely
reduced. 

\begin{figure}
\def\tmpwidth{0.6\textwidth}
\centering
% \begin{adjustbox}{width=\textwidth}
%\resizebox{\textwidth}{!}{%
\begin{tikzpicture}[every node/.style={inner sep=0,outer sep=0,inner xsep=0pt,outer xsep=0pt}]
\begin{axis}[%
width=0.9\textwidth,
axis x line=top,
axis y line=center,
x axis line style={draw=none},
xtick=\empty,
ymin=-42,ymax=32.4,%2.4,
xmin=-12,xmax=12,
  ytick={-37.5,29},
clip=false, %damit ich meine colorbar darunter bekomme
enlarge y limits=false,
yticklabels={0,32.4},
ylabel={$x$~[nm]},
y=-0.1cm, %y-vector, es ginge auch y={(xunit,yunit)}
x=0.5cm,
% ydir=reverse,
]
\node[anchor=north east,rotate=-90] at (-9.5,32){\includegraphics[width=\tmpwidth]{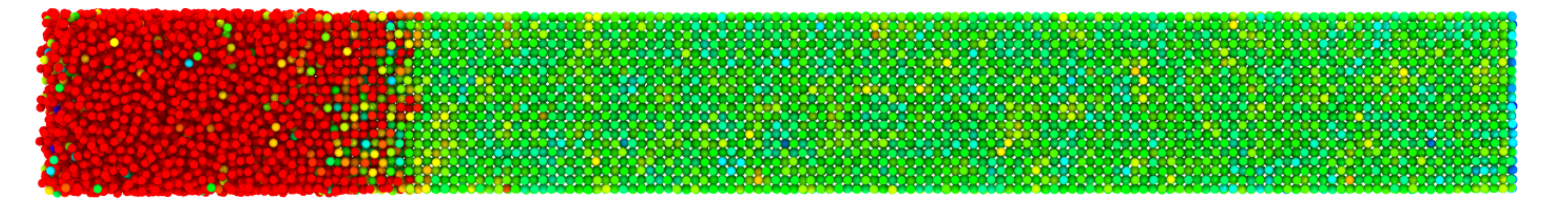}};
\node[anchor=north east,rotate=-90] at (-7.5,32) {\includegraphics[width=\tmpwidth]{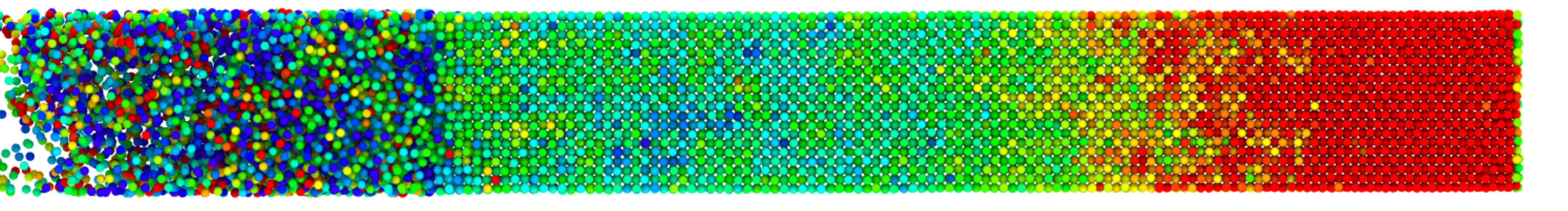}};
\node[anchor=north east,rotate=-90] at (-5.5,32) {\includegraphics[width=\tmpwidth]{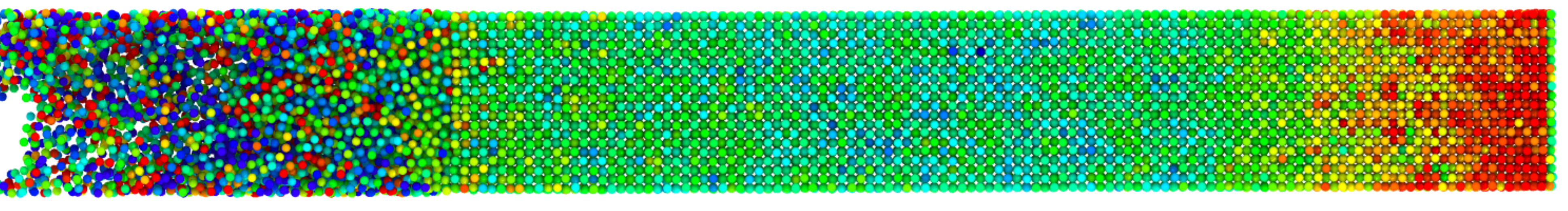}};
\node[anchor=north east,rotate=-90] at (-3.5,32) {\includegraphics[width=\tmpwidth]{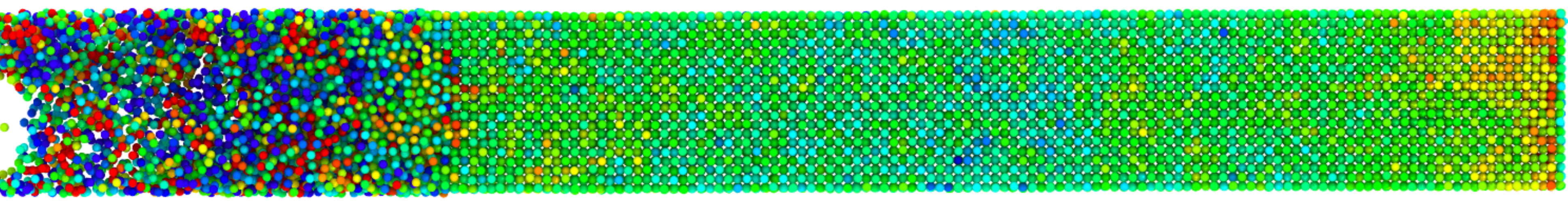}};
\node[anchor=north east,rotate=-90] at (-1.5,32) {\includegraphics[width=\tmpwidth]{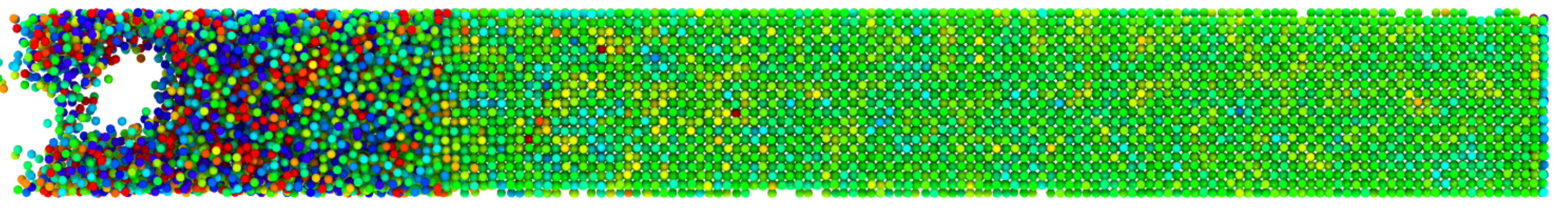}};
%RECHTE SEITE
\node[anchor=north east,rotate=-90] at (3.5,32){\includegraphics[width=\tmpwidth]{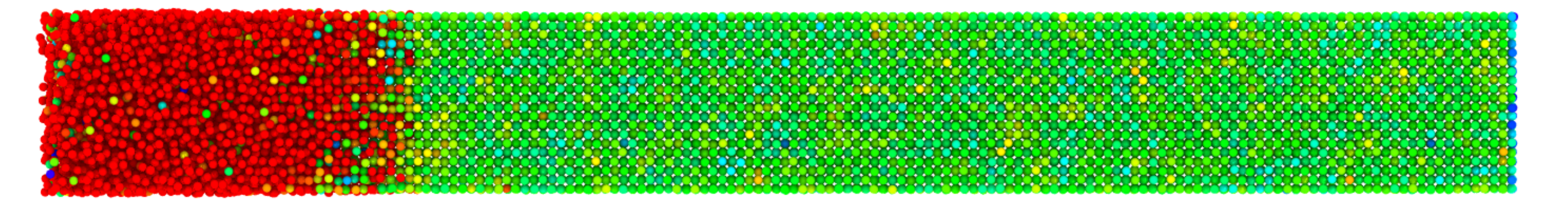}};
\node[anchor=north east,rotate=-90] at (5.5,32) {\includegraphics[width=\tmpwidth]{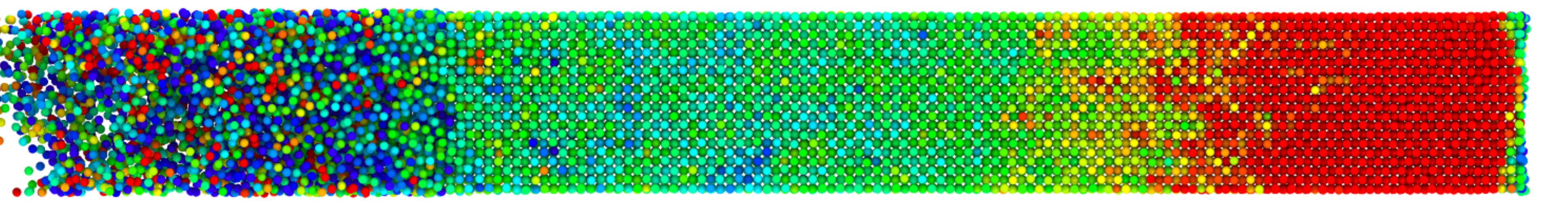}};
\node[anchor=north east,rotate=-90] at (7.5,32) {\includegraphics[width=\tmpwidth]{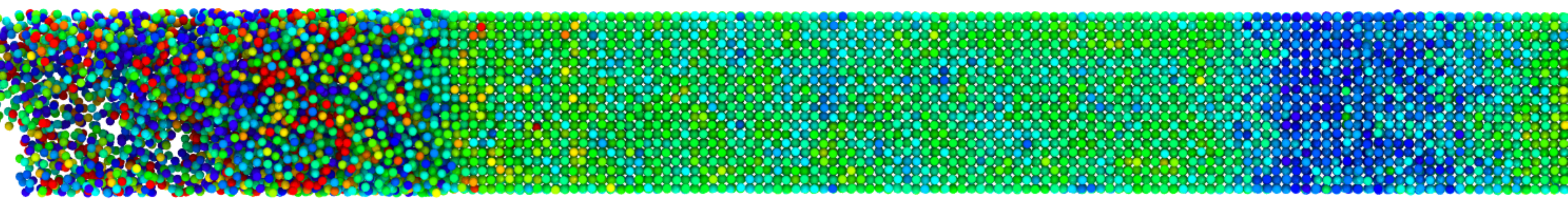}};
\node[anchor=north east,rotate=-90] at (9.5,32) {\includegraphics[width=\tmpwidth]{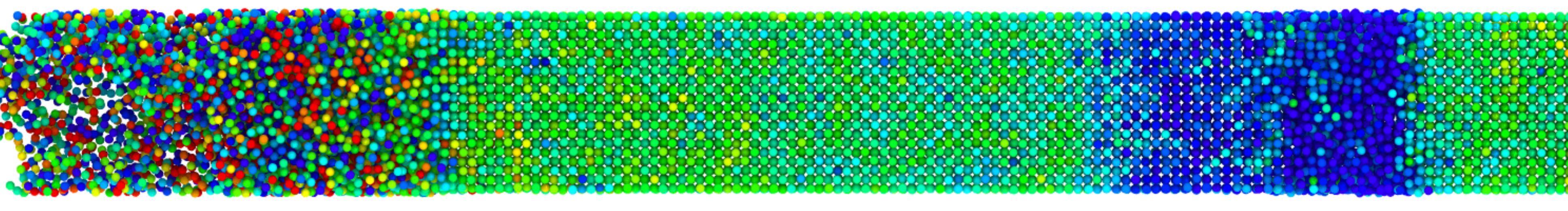}};
\node[anchor=north east,rotate=-90] at (11.5,32) {\includegraphics[width=\tmpwidth]{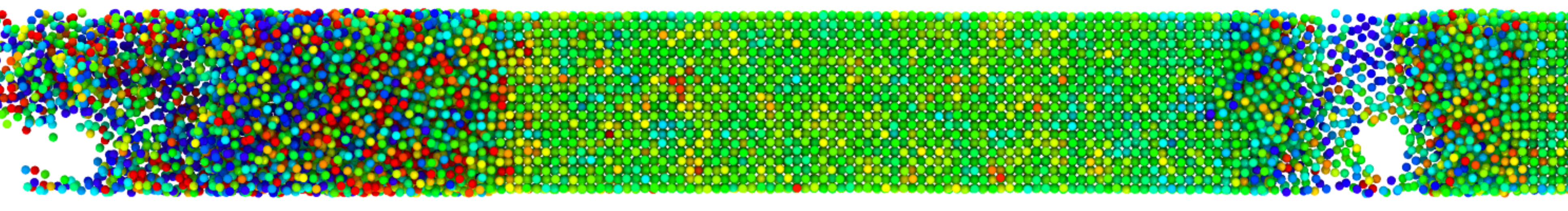}};
%COLORBAR
\node[anchor=north west](cbn) at (-3.5,34) {\includegraphics[width=0.3\textwidth]{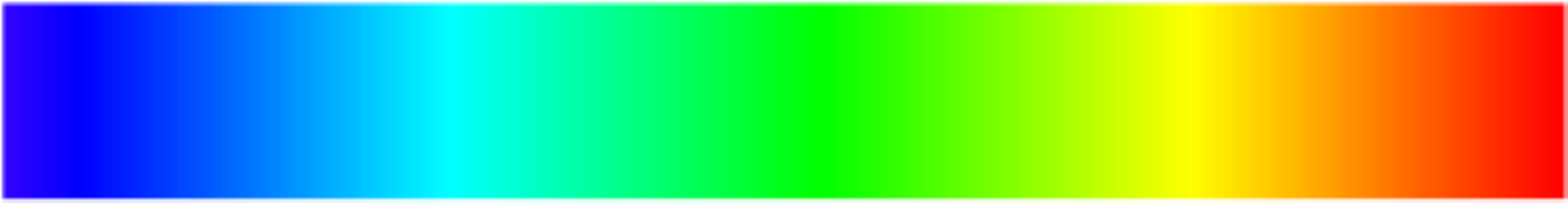}};
\node[below=0.2 cm of cbn]{pressure $\times$ atomic volume~[eV]};
\node[left=0.2 cm of cbn]{-0.9};
\node[right=0.2 cm of cbn]{0.9};
%ANNOTATIONS
\draw[->,line width=0.5mm] (-10.3,-23)--(-10.3,20);
\node[draw,fill=white,minimum size=1.4em] (abs) at (-4.5,-5) {absorption};
\draw[->,line width=0.5mm,below=0.1 cm of abs] (abs)--(-4.5,26);
\node[draw,fill=white,minimum size=1.4em] (refl) at (7,-5) {reflection};
\draw[->,line width=0.5mm,below=0.1 cm of abs] (refl)--(8.5,9);
\node[draw,fill=white,minimum size=1.4em] (spl) at (10.5,-5) {spallation};
\draw[->,line width=0.5mm,below=0.1 cm of abs] (spl)--(10.5,12);
%%ZEITPUNKTE
\node at (-10.5,-45) {\tiny 0.2~ps};
\node at (-8.5,-45) {\tiny  4~ps};
\node at (-6.5,-45) {\tiny  5~ps};
\node at (-4.5,-45) {\tiny  5.5~ps};
\node at (-2.5,-45) {\tiny  10~ps};
\node at (2.5,-45) {\tiny 0.2~ps};
\node at (4.5,-45) {\tiny  4~ps};
\node at (6.5,-45) {\tiny  5~ps};
\node at (8.5,-45) {\tiny  5.5~ps};
\node at (10.5,-45) {\tiny  10~ps};
\end{axis}
\end{tikzpicture}
%}
% \end{adjustbox}
\caption{Molecular dynamics snap shots of the hydrostatic stress after passage
  of shock and tension waves after sudden heating of the material surface at
  $x=0$. On the left side, the situation with pressure absorbing boundary conditions is displayed 
  and the right side represents the situation without any modifications of the original equations of motion. 
  The absorption of the pressure wave at the rear side of the sample can
  be seen clearly in the left part of the figure. The crystal structure is
  conserved despite the temporary buckling at the end of the sample. The same
  shock wave leads to spallation at the read side of the sample without pressure
  absorbing boundary conditions. In addition, the reflected stress 
  wave soon reaches the heat affected zone where it leads to an
  enlargement of the distorted region. Furthermore, several stress-nuclei are
  left behind, even below the heat affected zone. Along the directions 
  perpendicular to the beam direction, periodic boundary conditions have been applied.}
\label{fig:nrbatomic}
\end{figure}

The inspection of the atomistic snap shots in \ref{fig:nrbatomic} shows the
differences even more clearly. The wave at the rear side of the sample is gradually
absorbed and after 10~ps the sample is already nearly completely stress-free except for 
the heat affected zone (left). Without the modified equations
of motion it cannot be avoided that, disregarding the spallation, the stresses 
of the reflected wave further drive the ablation process of the material at the surface. 
Moreover, the passage of the reflected stress wave gives rise to numerous stress-nuclei 
even below the heat affected zone.

\section{Coupling to the TTM}

The pressure absorbing boundary conditions involved only the atomistic part of
the problem. In the combined MD+TTM model light might penetrate several
micrometers deep into the sample and heat up the sample. This would again
require very long samples which effectively represent a waste of resources
since nothing of interest is going on so deep into the sample. Replacing
atomistics by a finite-difference description can solve this problem.

\section{Implementation of the two temperature model}\label{sec:numttm}

The numerical treatment of the energy balance equation is
non-trivial and thus should be considered in some detail. The source term of
the irradiated laser energy and coupling term to the molecular dynamics
simulations will not be considered since they are irrelevant for the
description of the essential numerics. 
The balance equation for the internal, specifc energy of the electronic subsystem $e_e$ (internal energy per mass of the corresponding material), 
involving only drift and diffusion may be formulated as 
\begin{equation} 
 \frac{\partial (\rho e_e)}{\partial t} =\nabla\cdot(\kappa_e \nabla
 T_e)-\nabla\cdot (\vek{u}\rho e_e)\label{eq:ttmnum}\,,\\ 
\end{equation}
where $\kappa_e$ corresponds to the electronic heat conductivity, $T_e$ is the electronic temperature, $\rho$ represents the material density, and 
$\vek{u}$ is the drift velocity. 
%%%%%%%%%%%%%%%%%%%%%%%%%%%%%%%%%%%%%%%%%%%%%%%%%%%%%%%%%%%%%%%%%%%%%%%%

The first step to solve equation \ref{eq:ttmnum} is to chose a suitable
discretization of the simulation domain. 
Since molecular dynamics are combined with a hydrodynamical description of the electronic subsystem in the 
present hybrid approach, it is favorable to use the MD domain decomposition of
the sample into partial volumes also for the hydrodynamics part. 
Zhigilei and Ivanov\cite{Zhigilei2004} described how every MD-cell or typically a group
of such cells can be interpreted as a node of a finite-difference (FD)
grid. 
The volume of the FD node corresponds to the combined volumes of the
participating MD-cells.

\begin{figure}[!ht]
\centering
\begin{tikzpicture}
\begin{axis}[%
grid=none,
width=0.9\textwidth,
height=0.35\textwidth,
xlabel={$x$~[$\mathrm{\mu m}$]},
ymin=-0.1,ymax=1,
xmin=0.0,xmax=1.0,
legend style={at={(0.03,0.97)}, anchor=north west, legend cell align=left, align=left, draw=white!15!black},
x=12 cm,
y=9 cm,
hide axis,
]
\addplot graphics [xmin=0,xmax=1,ymin=0,ymax=1]{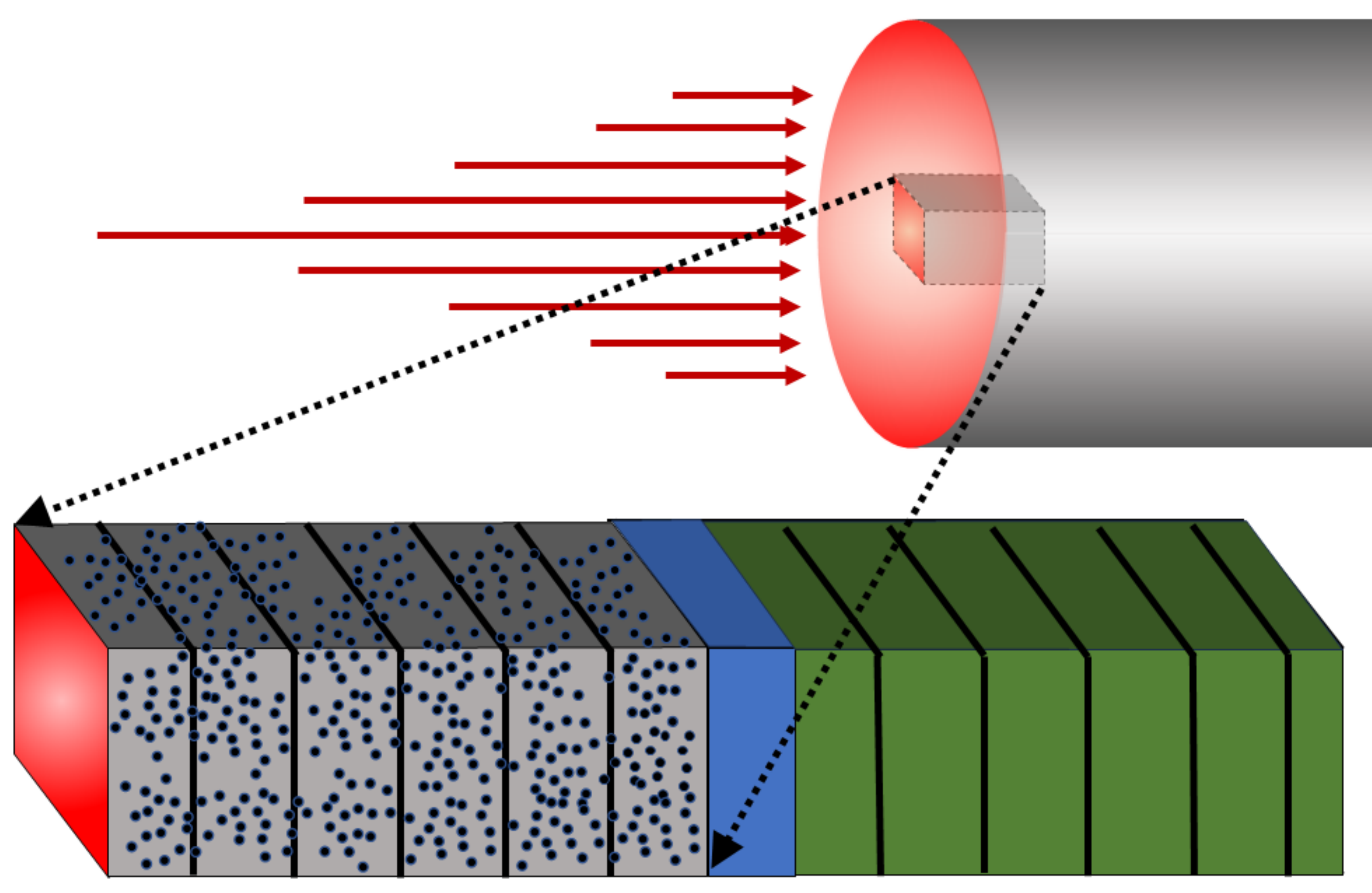};
%ANNOTATIONS
\draw [|-|] (0.08,0.0) -- node[below]{MD/TTM}(0.51,0.0);
\draw [|-|] (0.585,0.0) -- node[below]{TTM}(0.975,0.0);
\draw [|-|] (0.3,0.43) -- node[above]{$\Delta x$}(0.37,0.43);
\node (A) at (0.55,-0.08){pressure absorbing boundary conditions};
\draw[->] (A) -- (0.55,-0.01);
%\node[anchor=east] (la) at (0.1,0.83){$-0,5$};
\end{axis}
\end{tikzpicture}
\caption{Schematic illustration of the hybrid MD/TTM ansatz. The simulation focuses on a 
  small volume element of the whole sample, which is located in the very center of
  the laser beam. The lateral beam profile can be neglected in this
  one-dimensional approximation. The MD part of the model is limited to the
  heat affected zone to save resources. The MD simulation domain is enlarged
  by a ``virtual'' FD-lattice at the end of the sample to guarantee a
  natural heat flux of the electronic sub-system.} 
\label{fig:ttmhybrid}
\end{figure}
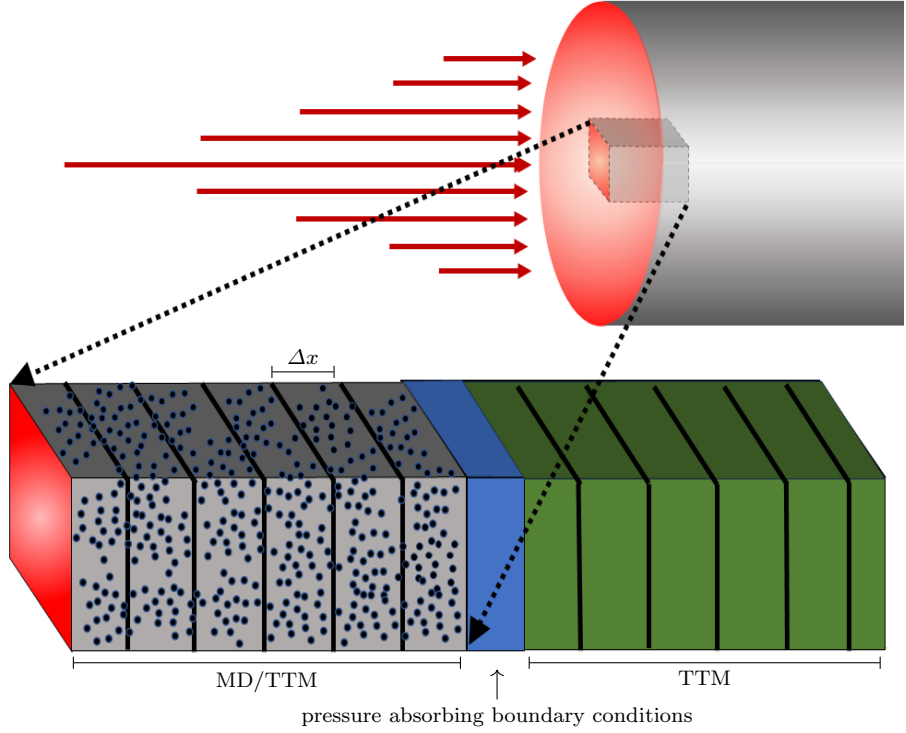

Typical beam diameters in experiments are in the range of a few microns. 
For metallic samples they are thus far larger than the optical penetration
depth of light which lies in the range of $10-20\,\mathrm{nm}$. 
For this reason it is sufficient to model the laser-metal interaction in a one-dimensional
approximation. 
By neglecting the transverse beam profile, the surface of the sample is heated homogeneously by the laser. 
The resulting temperature gradient thus exhibits, to a good approximation only a component parallel to 
the optical beam axis. 
The perpendicular components are negligibly small, even if the material is not isotropic. 
This permits to apply periodic boundary conditions along these transverse directions. 
In the end a tiny volume element of the sample in the center of the beam is considered. 
This situation is displayed in Fig.~\ref{fig:ttmhybrid}. 
To further reduce the computational load, an additional FD-lattice is attached at the end of the MD-simulation
domain, where the original two-temperature model is treated only on the continuum level. 
The MD-simulation domain is thus limited to the interesting part of the sample where the material is ablated, melts, dislocations and
other material restructuring processes occur. 
The task of the ``virtual'' FD-lattice only consists in providing a natural heat flux of the electronic subsystem.
Otherwise heat might build up at the rear side of the sample and tamper
the temperature gradient. 
Within the virtual lattice, heat diffusion is limited to the electronic subsystem only. 
The relatively slow heat diffusion of the lattice is neglected. 
The equation of state according to
Levashov\cite{Levashov2007} yields for the heat capacity of the lattice at
300~K a value of $8.59\cdot 10^2\,\mathrm{JK^{-1}kg^{-1}}$. 
The pressure absorbing boundary conditions are located between the two simulation domains
as shown in\ref{fig:ttmhybrid}. 
The modification of the equations of motion as they are exemplified in \ref{subsec:NRB} not only lead to the absorption of
elastic waves but reduce also the kinetic energy of the atoms and thus the temperature at
the rear side of the sample. 
For this reason, the virtual lattice does not couple directly to the end of the sample, but contrary to the situation shown in
figure, few atomic layers further down into the crystal lattice, where the lattice is not affected by the cooling effect of the boundary conditions. 
The region in between serves as a thermal buffer only.

An important question which has to be addressed first is how large such a
FD-cell has to be.
The question is in principle equivalent to the question if a scale separation
exists between the collision-mediated relaxation to the local equilibrium and
the diffusion-determined relaxation to the global equilibrium which permits a
continuum treatment of the particle collective. A detailed treatment of the
problem can be found in \cite{eisfeld2020}. Here we will only summarize the
results.

The mean square displacement of a particle $l^2$ has to be much larger than
the mean-free path between two collisions $\lambda_\text{mfp.}$.
The continuum description of the electron collective is indeed valid if 
\begin{equation}
 l\gg \lambda_\text{mfp.}\,.
\end{equation}

For aluminum the mean thermal velocity can be approximated in a large
temperature range by the Fermi-velocity $v_F\approx 1.5\cdot
10^6~\text{m/s}$. Furthermore the effective collision frequency is equal to
$\Gamma_\text{eff.}\approx 8.5\cdot 10^{14}$/s \cite{Eidmann2000} at room
temperature already, such that the lower limit of the mean free path
$\lambda_\text{mfp.}\approx 1.2~\text{fs}\cdot 2\cdot 10^6~\text{m/s}\approx
2.4\cdot 10^{-9}$~m. A FD cell should therefore possess a characteristic
length of a few nanometers to permit a continuum description of the enclosed
electron gas. At a temperature of $T=T_i=T_e=3000$~K for example
$\Gamma_\text{eff.}\approx 2.6\cdot 10^{16}$/s \cite{Fisher2002} is obtained
and the characteristic length results in $\approx 0.8$~nm only. $T_i$ is the
lattice temperature.

As important as the size of the FD cell is the number of the atoms contained
inside the cell, since in MD the temperature corresponds to the average kinetic
energy of an atom which moves with the most probable velocity
$v_\text{therm.}$ of the Maxwell-Boltzmann distribution.  

In order to be able to define a temperature for an atom collective, first a Maxwell-Boltzmann
distribution has to be present and second, the most probable velocity has to be
known. The first condition is problematic already since the strong
acceleration of the atoms by the heated electrons forces the lattice atoms
into a temporal non-equilibrium. This fact can be neglected if the equilibrium
is reached fast enough. Kinetic gas theory again provides a rough estimate
based on a simple geometric consideration of two elastically colliding hard
spheres. If the deviation from the Maxwell-Boltzmann distribution is small
then the relaxation time is calculated with \cite{kaviany2014}
\begin{equation}
 \tau_\text{MB}=\frac{m^{1/2}}{4\pi^{1/2}d^2(k_B T)^{1/2} n_i}\,,
\end{equation}
where for aluminum, the mass of the atom $m\approx 4.32\cdot 10^{-26}$~kg,
the particle density at standard conditions $n_i\approx 6.25\cdot
10^{28}$~$\text{m}^{-3}$ and the particle diameter, as well as the average distance
resulting from the particle density, respectively, $d\approx 1.56\cdot
10^{-10}$~m. Thus at $T=300$~K the relaxation time
$\tau_\text{MB}(300\,\text{K})\approx 3\cdot 10^{-13}$~s and at high
temperatures, for example $\tau_\text{MB}(10000\,\text{K})\approx 3.5\cdot
10^{-14}$~s. 
Therefore, for minor deviations from the equilibrium distribution, the actual non-equilibrium may be 
neglected for increasing temperatures.

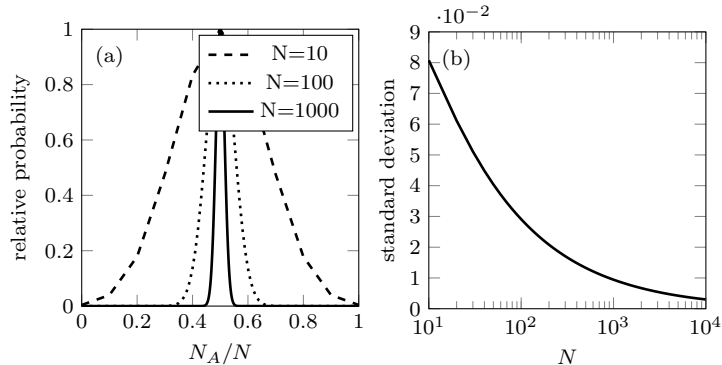
\begin{figure}[!h]
\centering
\begin{tikzpicture}
\begin{axis}[%
width=0.45\textwidth,
height=0.45\textwidth,
grid=none,
%at={(0,0)},
%scale only axis,
xmin=0,
xmax=1,
xlabel={$N_A/N$},
ymin=0,
ymax=1,
ylabel={relative probability},
]
\addplot
	 [
	   color=black,dashed,  
	   line width=1.0pt	  
	 ]
	 table {plot/data/omega_vs_N10.dat};
\addlegendentry{N=10}

\addplot
	 [
	   color=black, dotted, 
	   line width=1.0pt	  
	 ]
	 table {plot/data/omega_vs_N100.dat};
\addlegendentry{N=100}

\addplot
	 [
	   color=black, 
	   line width=1.0pt	  
	 ]
	 table {plot/data/omega_vs_N1000.dat};
\addlegendentry{N=1000}
\node (foo) at (rel axis cs:0.1,0.9){(a)};
\end{axis}
\end{tikzpicture}%
\begin{tikzpicture}
\begin{axis}[%
width=0.45\textwidth,
height=0.45\textwidth,
grid=none,
%at={(0,0)},
%scale only axis,
xmode=log,
xmin=10,
xmax=10000,
xminorticks=true,
ytick distance=0.01,
xlabel={$N$},
ymin=0,
ymax=0.09,
ylabel={standard deviation},
axis background/.style={fill=white}]
\addplot
	 [
	   color=black, 
	   line width=1.0pt
	 ]
	 table {plot/data/std_vs_N.dat};
\node (foo) at (rel axis cs:0.1,0.9){(b)};	 
\end{axis}
\end{tikzpicture}%
\caption{Illustration of the relative probability to draw $N_A$ particles of
  type $A$ out of a reservoir of a total of $N$ equidistributed particles (a)
  and the corresponding standard deviation (b).}
\label{fig:binomdistrib}
\end{figure}

The second condition requires the number of atoms within a single FD cell to be big enough, 
such that the average atomic velocity exhibits a statistical significance.
Universally accepted is a number on the order of the Avogadro-constant $\approx
10^{23}$, while some authors refer to a few hundreds of atoms\cite{allen1989}. 
In this regard, a thought experiment might be helpful:

From a reservoir of infinitely many particles of type $A$ and $B$ subsequently
$N$ particles are drawn. The number of possibilities to draw $N_A$ particles
of type $A$ and $N_B$ particles of type $B$ is thus
\begin{equation}
 \Omega(N_A,N_B)=\frac{N!}{N_A!N_B!}=\frac{N!}{N_A!(N-N_A)!}=\begin{pmatrix} N
 \\ N_A\end{pmatrix}\,. 
\end{equation}
If the reservoir contains equally many particles of type $A$ and $B$ then the
most probable outcome of the experiment is $N_A=N_B=N/2$, where in the limit
\begin{equation}
 \lim\limits_{N\to\infty}\Omega(N_A,N_B) =\Omega(N/2,N/2)=\frac{N!}{(N/2)!(N/2)!}\,.
\end{equation}
If the resulting distribution is presented as a function of $N_A/N$ or
$N_B/N$ then, with increasing $N$ the distribution converges to a Gaussian
centered at $1/2$ with a standard deviation $\propto N^{-1/2}$, see Fig. \ref{fig:binomdistrib}.  

This simple thought experiment can be generalized to an arbitrary number of particle
types which, in a transferred sense, can represent separate energy levels (in the
NVE ensemble). 

In this case, the $x$-axis corresponds to the energy of the state in
consideration $E_i$ and the $y$-axis to the occupation probability
$P(E_i)=\Omega(E-E_i)/\sum_j \Omega(E-E_j)$ which would be maximal at the
internal energy $U=\langle E\rangle$, where $E$ represents the total energy of the
system.

The standard deviation for $N\approx 100$ results in about 3\% while for
$N\approx 1000$ it falls below 1\% as can be seen in Fig. \ref{fig:binomdistrib}(b).
This confirms the claim in \cite{allen1989} that a few hundred atoms lead to a
reasonable average of the thermal energy already and thus provide a useful
estimation of the temperature.

\section{Conclusion}

Two improvements for large-scale computer simulations have been described. In
materials science there are many studies of multi-scale simulations coupling
atomistic and meso-scale finite-element methods. Sometimes even a further
layer is added describing the interaction by ab-initio-methods. These
approaches typically work only due to the fact that they are close to
equilibrium and no dynamical processes like strong waves occur. Here we have
described on how to erase strong waves and how to consistently deal with
long-range phenomena avoiding expensive atomistic simulations. The hope is
that the ideas presented can also be applied in other circumstances.

\acknowledgement
Eugen Eisfeld has been supported by the DFG within subproject B.5 in the
formers SFB 716. Dominic Klein is supported by the Hans-B\"{o}ckler-Stiftung.

%%%%%%%%%%%%%%%%%%%%%%%%%%%%%%%%%%%%%%%%%%%%%%%%
%% BACKMATTER
%%%%%%%%%%%%%%%%%%%%%%%%%%%%%%%%%%%%%%%%%%%%%%%%

%\bibliographystyle{aipproc}   % if natbib is available
%\bibliographystyle{aipprocl} % if natbib is missing

%%%%%%%%%%%%%%%%%%    Bib    %%%%%%%%%%%%%%%%%%%%%%%%%%
%\printbibliography


\begin{thebibliography}{}

\bibitem{allen1989}Allen, M. \& Tildesley, D. Computer simulation of liquids. (Oxford university press,1989)
\bibitem{Eidmann2000}Eidmann, K., Meyer-Ter-Vehn, J., Schlegel, T. \& HÃ¼ller, S. Hydrodynamic simulation of subpicosecond laser interaction with solid-density matter. {\em Physical Review E - Statistical Physics, Plasmas, Fluids, And Related Interdisciplinary Topics}. \textbf{62}, 1202-1214 (2000)
\bibitem{eisfeld2020}Eisfeld, E. Molekulardynamische Simulationen der Laserablation an Aluminium unter Einbeziehung von Plasmaeffekten. (Universität Stuttgart,2020)
\bibitem{Fang2012}Fang, M., Tang, S., Li, Z. \& Wang, X. Artificial boundary conditions for atomic simulations of face-centered-cubic lattice. {\em Computational Mechanics}. \textbf{50}, 645-655 (2012)
\bibitem{Fisher2002}Fisher, D., Fraenkel, M., Henis, Z., Moshe, E. \& Eliezer, S. Interband and intraband (Drude) contributions to femtosecond laser absorption in aluminum. {\em Physical Review E - Statistical Physics, Plasmas, Fluids, And Related Interdisciplinary Topics}. \textbf{65}, 1-8 (2002)
\bibitem{gumbsch}Gumbsch, P., Zhou, S. \& Holian, B. Molecular dynamics investigation of dynamic crack stability. {\em Phys. Rev. B}. \textbf{55}, 3445-3455 (1997,2), https://link.aps.org/doi/10.1103/PhysRevB.55.3445
\bibitem{Karpov2006}Karpov, E., Yu, H., Park, H., Liu, W., Wang, Q. \& Qian, D. Multiscale boundary conditions in crystalline solids: Theory and application to nanoindentation. {\em International Journal Of Solids And Structures}. \textbf{43}, 6359-6379 (2006)
\bibitem{kaviany2014}Kaviany, M. Heat transfer physics. (Cambridge University Press,2014)
\bibitem{klein2021}Klein, D. Simulation of Laser Ablation in Covalent Materials, in preparation. (Universität Stuttgart)
\bibitem{Levashov2007}Levashov, P. \& Khishchenko, K. Tabular multiphase equations of state for metals and their applications. {\em AIP Conference Proceedings}. \textbf{955} pp. 59-62 (2007)
\bibitem{ming2009}Ming, F. \& Shao-Qiang, T. Efficient and robust design for absorbing boundary conditions in atomistic computations. {\em Chinese Physics Letters}. \textbf{26}, 116201 (2009)
\bibitem{pang2011}Pang, G. \& Tang, S. Time history kernel functions for square lattice. {\em Computational Mechanics}. \textbf{48}, 699-711 (2011)
\bibitem{Park2005}Park, H., Karpov, E. \& Liu, W. Non-reflecting boundary conditions for atomistic, continuum and coupled atomistic/continuum simulations. {\em International Journal For Numerical Methods In Engineering}. \textbf{64}, 237-259 (2005)
\bibitem{md-ttm}Roth, J., Trichet, C., Trebin, H. \& Sonntag, S. Laser
  Ablation of Metals. {\em High Performance Computing In Science And
    Engineering '10}. pp. 159 (2011)
\bibitem{Schafer2002}Schäfer, C., Urbassek, H., Zhigilei, L. \& Garrison, B. Pressure-transmitting boundary conditions for molecular-dynamics simulations. {\em Computational Materials Science}. \textbf{24}, 421-429 (2002)
\bibitem{DissSonntag}Sonntag, S. Computer Simulations of Laser Ablation from Simple Metals to Complex Metallic Alloys. (Universität Stuttgart,2010)
\bibitem{ulrich}Ulrich, C. Simulation der Laserablation an Metallen, Diploma Thesis. (Universität Stuttgart,2007)
\bibitem{Zhigilei2004}Zhigilei, L., Ivanov, D., Leveugle, E., Sadigh, B. \& Bringa, E. Computer modeling of laser melting and spallation of metal targets.  (2004)
\end{thebibliography}
\end{document}